\documentclass[conference]{IEEEtran}
\IEEEoverridecommandlockouts

\makeatletter
\def\markboth#1#2{\def\leftmark{\@IEEEcompsoconly{\sffamily}\MakeUppercase{\protect#1}}%
\def\rightmark{\@IEEEcompsoconly{\sffamily}\MakeUppercase{\protect#2}}}
\makeatother

\newcommand\blfootnote[1]{%
  \begingroup
  \renewcommand\thefootnote{}\footnote{#1}%
  \addtocounter{footnote}{-1}%
  \endgroup
}
\usepackage{amsmath}

\usepackage[english]{babel}
\usepackage{xcolor}
\usepackage{lipsum}
\usepackage{caption}
\usepackage{cite}
\usepackage[pdftex]{graphicx}
\usepackage{subcaption}
\usepackage{amsmath}
\usepackage{booktabs}
\usepackage{colortbl}
\usepackage{amsfonts}
\usepackage{amssymb}
\usepackage{amsthm}
\usepackage{array}
\usepackage{verbatim}
\usepackage{listings}
\usepackage{algorithm}
\usepackage{algpseudocode}
\usepackage{url}
\usepackage{enumerate}
\usepackage{multirow}

\definecolor{LightBlue}{rgb}{0.5,0.5,1}
\definecolor{LightRed}{rgb}{1,0.5,0.5}
\definecolor{LightYellow}{rgb}{1,0.85,0}

\usepackage{epsfig}
\usepackage{epstopdf}
\usepackage{multicol}
\usepackage[font=footnotesize]{caption}

\usepackage{hyperref}
\usepackage{amsmath}
\usepackage{algorithm}

\makeatletter
\def\BState{\State\hskip-\ALG@thistlm}
\makeatother

\DeclareMathOperator*{\argmax}{arg\,max}

\newcommand{\bi}{\begin{itemize}}
\newcommand{\ei}{\end{itemize}}
\newcommand{\be}{\begin{equation}}
\newcommand{\ee}{\end{equation}}

\def\beq{\begin{equation}}
\def\eeq{\end{equation}}
\def\beqa{\begin{eqnarray}}
\def\eeqa{\end{eqnarray}}
\def\beqan{\begin{eqnarray*}}
\def\eeqan{\end{eqnarray*}}

\title{Hybrid Spectrum Access for mmWave Networks}
\author{{{\bf Mattia Rebato}$^\dagger$, {\textbf{Federico Boccardi}$^\diamond$}, {\bf Marco Mezzavilla}$^*$, {\bf Sundeep Rangan}$^*$, {\bf Michele Zorzi}$^\dagger$ }\\

$^\dagger$ University of Padova, Italy\qquad\qquad$^*$ NYU WIRELESS, Brooklyn, NY, USA  \\
\small{$\{$\texttt{rebatoma}, \texttt{zorzi}$\}$\texttt{@dei.unipd.it}, \texttt{federico.boccardi@ieee.org} $\{$\texttt{mezzavilla}, \texttt{srangan}$\}$\texttt{@nyu.edu},
}
}

\begin{document}
\maketitle
\begin{abstract}
While spectrum at millimeter wave (mmWave) frequencies is less scarce than at traditional frequencies below 6~GHz, still it is not unlimited, in particular if we consider the requirements from other services using the same band and the need to license mmWave bands to multiple mobile operators. Therefore, an efficient spectrum access scheme is critical to harvest the maximum benefit from emerging mmWave technologies. In this paper, motivated by previous results where spectrum pooling was proved to be more feasible at high mmWave frequencies, we  study the performance of a hybrid spectrum scheme where exclusive access is used at frequencies in the 20/30~GHz range while spectrum pooling/unlicensed spectrum is used at frequencies around 70~GHz.
Our preliminary results show that hybrid spectrum access is a promising approach for mmWave networks, and motivate further studies to achieve a more complete understanding of both technical and non technical implications.

\blfootnote{$^\diamond$ F. Boccardi's work was carried out in his personal capacity and the views expressed here are his own and do not reflect those of his employer.}
\end{abstract}
\bigskip

\section{Introduction}
\label{introduction}
The potential of orders of magnitude increases in capacity offered by the millimeter wave (mmWave) frequencies poses new spectrum access challenges. In this paper, we introduce a hybrid scheme that builds on both exclusive access and spectrum pooling concepts.

Traditionally, wireless data services have been delivered mainly by using two different spectrum access models. Under the \textit{exclusive} model, each mobile operator is granted the right of exclusive use of a spectrum band to provide mobile services. Exclusive spectrum access has been one of the key factors for the successful deployment of cellular systems since their inception, and it is by far the default model to provide mobile services. Under the \textit{license-exempt} (also referred to as  \textit{unlicensed}) model, spectrum is allowed to be used by several users/mobile operators. While there is no guaranteed access to an instantaneously fixed amount of spectrum, politeness rules (e.g., based on a listen-before-talk principle) are in place to allow a fair use of the spectrum. The license-exempt spectrum model has been one of the  key factors for the successful deployment of WiFi as a ubiquitous way of connecting devices to the Internet. The \textit{spectrum pooling} model has also been considered as an intermediate paradigm, where different operators are granted access to the same spectrum resources, with rules that are known a priori. Spectrum pooling does not provide guarantees for the access to an instantaneously fixed amount spectrum, but ensures some level of predictability and short-term and long-term fairness~\cite{METIS51}. 
\IEEEpubidadjcol

Recently, new technologies emerged that aggregate spectrum in both exclusive and license-exempt bands, in a way to route the different information pipes to the carrier that best matches their requirements. Aggregation can be implemented at the MAC layer, to allow a very rapid switch between exclusive and license-exempt carriers, effectively realising a  \textit{hybrid spectrum access regime}.
Examples of these technologies are Licensed Assisted Access (LAA), LTE-WiFi link aggregation and LTE-WiFi interworking.
We note that hybrid spectrum access has been proposed for traditional  spectrum below 6~GHz as an extension of carrier aggregation, while to the best of our knowledge there are no works assessing the benefit of hybrid spectrum access for mmWave networks.


While spectrum at mmWave frequencies is less scarce than at traditional frequencies below 6~GHz, still it is not unlimited, in particular if we consider the requirements from other services (e.g., satellite and fixed services) and the need to license mmWave bands to multiple mobile operators. Therefore, an efficient spectrum access scheme is critical to harvest the maximum benefit from emerging mmWave technologies ~\cite{boccardi16}. Recent works compared exclusive spectrum allocation with different types of spectrum pooling or unlicensed models, showing different results as a function of the assumptions used. Reference~\cite{li2014} introduced a new signaling report among mobile operators, to establish an interference database to support scheduling decisions, with both a centralized and a distributed supporting architecture. In the centralized case, a new architectural entity receives information about the interference measured by each network and determines which links cannot be scheduled simultaneously. In the decentralized case, the victim network sends a message to the interfering network with a proposed coordination pattern. The two networks can further refine the coordination pattern via multiple stages. Reference~\cite{gupta15} studied the feasibility of spectrum pooling in mmWave networks under the assumption of ideal antenna patterns and showed that spectrum pooling might be beneficial even without any coordination between the different operators. In particular,~\cite{gupta15} showed that uncoordinated pooling provides gain at both 28~GHz and 73~GHz.  Reference~\cite{boccardi16} further developed the results in~\cite{li2014} and~\cite{gupta15},  focusing on the effect of coordination and of inaccurate beamforming. It showed that, while coordination may not be needed under ideal assumptions, it does provide substantial gains when considering more realistic channel and interference models and antenna patterns. Moreover,  it showed that, under realistic assumptions, spectrum pooling without coordination might be more feasible at high mmWave frequencies (e.g., 70~GHz) than at low mmWave frequencies (e.g., 28 or 32~GHz), due to the higher directionality of the beams. 
Reference~\cite{rebato16} compares different resource sharing paradigms and shows that a full spectrum and infrastructure sharing configuration provides significant advantages, even without resorting to complex signaling protocols for the exchange of information between multiple operators' networks.

\begin{figure}[h!]
\centering
\includegraphics[width=0.8\columnwidth]{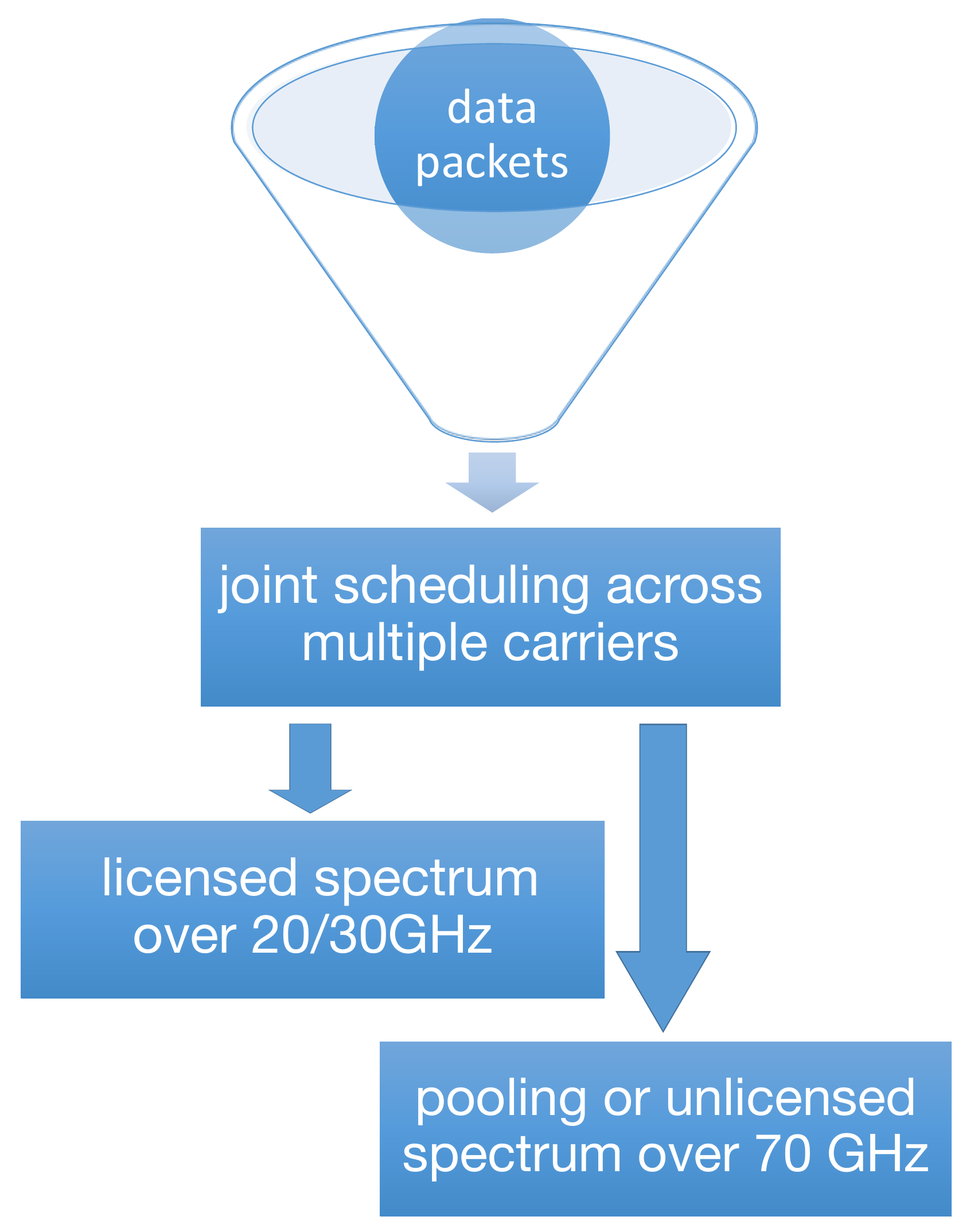}     
\caption{Blocks diagram of the joint scheduling that allocates user packets over the different bands.}
\label{block_scheme}
\end{figure}

This paper extends the previous results in~\cite{boccardi16} and~\cite{rebato16} to the case of hybrid spectrum allocation.
In other words, differently from the previous works, where exclusive access and spectrum pooling were compared, in this work we propose a spectrum access paradigm that builds on both exclusive access and unlicensed access/spectrum pooling\footnote{Note that our results apply to both unlicensed access and spectrum pooling for the mmWave carrier at higher frequency, as we consider an unplanned deployment of base stations with random positions. Such a random deployment is reminiscent of the case where BSs are not deployed by operators via a careful planning but by final users. However, for the case of unlicensed access further work is needed to study politeness protocols that can guarantee a fair use of the spectrum to the different spectrum users.}.
In particular, motivated by the results in~\cite{boccardi16} where pooling was proved to be more feasible at high mmWave frequencies, we  study the performance of a hybrid spectrum scheme where exclusive access is used at frequencies in the 20/30~GHz range while pooling is used at frequencies around 70~GHz\footnote{In the following we will refer to the 28~GHz and 73~GHz bands, for which many measurements are available in the literature (e.g., see~\cite{mustafa,ted1,ted2,ted3,ted4}). However, we note that the results herein, possibly with same minor modification, would apply to adjacent bands as well. In particular the results obtained for the 28~GHz band apply also to the two bands selected by the 2015 World Radio Conference (WRC-15) for sharing and compatibility studies for 5G, i.e., 24.25 - 27~GHz and 31.8 - 33.5~GHz. The results obtained for 73~GHz apply to the 66 - 76~GHz band, again selected by WRC-15 for sharing and compatibility studies for 5G.}. The two bands are aggregated at the MAC layer as in Figure~\ref{block_scheme}, and users are allocated to one or the other band to maximize the rate. In this way, interference-limited users are allocated to the exclusive spectrum component while noise-limited users are allocated to the pooled spectrum components. We compare our proposal with two baselines, one relying on exclusive spectrum access at both 28~GHz and 73~GHz and thus favoring interference-limited scenarios, the other relying on pooling at both 28~GHz and 73~GHz and thus favoring noise-limited scenarios. Our preliminary results show that hybrid spectrum access is a promising approach for mmWave networks, and motivate further studies for a more complete understanding of both technical and non technical implications.

The rest of the paper is organized as follows. In Section \ref{system_model}, we describe the system model. In Section \ref{Hybrid_UE_association}, we describe the proposed hybrid spectrum allocation algorithm. In Section \ref{simulation_results}, we provide a numerical evaluation and discuss the results. Finally,  we conclude the paper and describe some future research steps in Section \ref{conclusion}.

\section{System model}
\label{system_model}
We consider a scenario with $M$ operators, each with a set $\mathcal{I}_m$ of base stations (BSs) and a set $\mathcal{J}_m$ of user equipments (UEs). For each operator, both UEs and BSs are placed following a Poisson point process (PPP) with densities $\lambda_{\text{UE}}$ and $\lambda_{\text{BS}}$, respectively, which represents an unplanned deployment, where base stations are not optimally located.
Moreover, we assume that all BSs support two mmWave bands, one at 28~GHz and one at 73~GHz. We assume a downlink transmission where for a given operator $ m \in \{1,\ldots,M\}$ we define $\mathbf{x}_{mij}$ as the $n_{\text{TX}} \times 1$ transmitted vector between BS $i \in \{1,\ldots,I_m\}$ and UE $j \in \{1,\ldots,J_m\}$\footnote{Note that $\mathcal{J}_m$ represents the set of users for operator $m$, while the term $J_m$ stands for its cardinality, so $J_m = |\mathcal{J}_m| \, \forall m \in \mathcal{M}$. The same concept is used also for operators and BSs, thus $I_m = | \mathcal{I}_m|$ and $M = |\mathcal{M}|$.}:
\begin{equation}
\mathbf{x}_{mij}=\mathbf{w}_{{\text{TX}}_{mij}}u_{mij},
\end{equation}
where $n_{\text{TX}}$ is the number of antennas at the BS transmitter, $\mathbf{w}_{{\text{TX}}_{mij}}$ is the $n_{\text{TX}} \times 1$ beamforming vector between the $i$-th BS and the $j$-th UE of the $m$-th operator, and $u_{mij}$ is the corresponding scalar information symbol transmitted.
We model the transmit antennas as a uniform planar array (UPA) at both BS and UE, and allow more antenna elements to be deployed at 73~GHz compared to 28~GHz.
In particular, we assume that the numbers of antennas at the BS transmitter and at the UE receiver are $n_{\text{TX}} = 64$ and  $n_{\text{RX}} = 16$ for the 28~GHz links, and $n_{\text{TX}} = 256$ and  $n_{\text{RX}} = 64$ for the 73~GHz links\footnote{Note that, thanks to the reduced wavelength, the bigger array at 73~GHz will have a comparable (in fact, slightly smaller) physical size than that at 28~GHz.}.
The beamforming gain from BS $i$ to UE $j$ of operator $m$ is given by:
\begin{equation}
G_{mij} = |\textbf{w}^T_{\text{RX}_{mij}} \textbf{H}_{mij} \textbf{w}_{\text{TX}_{mij}} |^2,
\label{bf_comp}
\end{equation}
where $\textbf{w}_{\text{TX}_{mij}} \in \mathbb{C}^{n_{\text{TX}}}$ is the beamforming vector of transmitter $i$ when transmitting to receiver $j$, $\textbf{w}_{\text{RX}_{mij}} \in \mathbb{C}^{n_{\text{RX}}}$ is the beamforming vector of receiver $j$ when receiving from transmitter $i$, and $\textbf{H}_{mij}$ is the channel matrix (described in the following).
Beamforming vectors are computed as follow:
\begin{equation}
\textbf{w}(\Omega) = \frac{1}{\sqrt {n}} \begin{bmatrix} 1\\ \exp(-j2\pi \Delta \Omega)\\ \exp(-j2\pi 2\Delta \Omega) \\ \vdots \\ \exp(-j2\pi (n-1) \Delta \Omega) \end{bmatrix}, 
\end{equation}
where $\Delta$ represents the spacing between the elements of the array, $\Omega$ is the direction of transmission or reception (considering both horizontal and vertical angles), and $n$ is the normalization factor and corresponds to the number of elements in the antenna array. 
We assume to have the possibility of steering in any direction.
Therefore, we can generate a beamforming vector for any possible angle between 0 and 360 degrees. We also assume perfect alignment between the beams of each UE and its serving BS.
We consider the following transmit power constraint: 
\begin{equation}
E\left[ \vert u_{mij}^{*} u_{mij}\vert ^2\right]  \leq P,
\end{equation}
where $P$ is the maximum power allowed at the input of the antenna array (or equivalently at the output of the power amplifier driving the antenna array).
The received signal at the $i$-th user of the $m$-th operator is modeled as follows:
\begin{equation}
 \textbf{y}_{mij} =
\textbf{H}_{mij}\textbf{x}_{mij} + \textbf{n}_{mij},
 \end{equation}
where $\textbf{H}_{mij}$ is the channel matrix\footnote{The channel is modeled for a dense urban environment as in~\cite{mustafa} and~\cite{mathew}.}, modeled as composed of a random number $K$ of clusters, each corresponding to a macro-level scattering path, and $\textbf{n}_{mij}$ is the additive noise (which also includes interference).
At the receiver, the number of clusters is given as the maximum between 1 and a Poisson random variable whose mean $\lambda$ is related to the carrier frequency as explained in \cite{mustafa}.
For each cluster $k$, the number of sub-paths is modeled as an integer random variable uniformly distributed in $\{1,\ldots, 10\}$.
Given a set of clusters and of sub-paths for a channel, we can compute the channel matrix as\footnote{In the rest of this section we remove the subscripts $mij$ for ease of notation.}:
\begin{equation}
\textbf{H}(t,f)=\sum_{k=1}^{K}\sum_{\ell=1}^{L_k}g_{k\ell}(t,f) \textbf{u}_{\text{RX}}(\theta^{\text{RX}}_{k\ell},\phi^{\text{RX}}_{k\ell}) \textbf{u}^*_{\text{TX}}(\theta^{\text{TX}}_{k\ell},\phi^{\text{TX}}_{k\ell}),
\label{channel_matrix}
\end{equation}
where $L_k$ is the number of sub-paths in cluster $k$, $g_{k\ell}(t,f)$ is the small-scale fading of path $\ell$ in cluster $k$ over time and frequency, and $\textbf{u}_{\text{RX}}(\cdot)$ and $\textbf{u}_{\text{TX}}(\cdot)$ are the spatial signature vectors of the receiver and the transmitter, respectively.
Spatial signatures are computed with horizontal and vertical angles of arrival (AoA) $\theta^{\text{RX}}_{k\ell},\phi^{\text{RX}}_{k\ell}$, and horizontal and vertical angles of departure (AoD) $\theta^{\text{TX}}_{k\ell},\phi^{\text{TX}}_{k\ell}$, where $k = 1, \dots, K$ is the cluster index and $\ell = 1, \dots, L_k$ is the sub-path index within the cluster.
The small-scale fading (obtained from~\cite{mustafa}) is generated based on the number of clusters, the number of sub-paths in each cluster, the Doppler shift, the power spread, and the delay spread, as:
\begin{equation}
g_{k\ell}(t,f)=\sqrt{P_{k\ell}}e^{j 2\pi f_{d}cos(\omega_{k\ell} )t}e^{-j2\pi \tau _{k\ell}f},
\label{scale_fading}
\end{equation}
where $P_{k\ell}$ is the power spread of sub-path $\ell$ in cluster $k$, $f_{d}$ is the maximum Doppler shift, $\omega_{k\ell}$ is the angle of arrival of sub-path $\ell$ in cluster $k$ with respect to the direction of motion, $\tau _{k\ell}$ is the delay spread of sub-path $\ell$ in cluster $k$, and $f$ is the carrier frequency.

The power spread $P_{k\ell}$ is obtained by following~\cite{mathew}:
\be
P_{k\ell} = \frac{P_{k\ell}^\prime}{\sum_{h=1}^K \sum_{p=1}^{L_h} P_{hp}^\prime}, \quad
P_{k\ell}^\prime = \frac{U_k^{r_{\tau}-1}10^{-0.1 Z_k+V_{k\ell}}}{L_k},
\ee
where $U_k \sim U[0,1]$, $V_{k\ell} \sim U[0,0.6]$ and $Z_k \sim N(0,\zeta^2)$, while parameters $r_\tau$  and $\zeta$ are found in~\cite{mustafa}.

Following the measurement campaign carried out in a real dense urban environment and reported in~\cite{ted1,ted2,ted3,ted4}, pathloss can be modeled with three states: line-of-sight (LoS), non-line-of-sight (NLoS) and outage. Each link is characterized by the channel state probabilities $p_{\mathrm{LoS}}$, $p_{\mathrm{NLoS}}$ and $p_{\mathrm{out}}$, which are expressed in terms of the distance $d$ between UE and BS as follows:
\be
\begin{matrix}
\vspace{-0.3cm}
p_{\mathrm{out}}(d)=\mathrm{max}(0,1-e^{-a_{\mathrm{out}}d+b_{\mathrm{out}}})\\ \vspace{-0.3cm}
p_{\mathrm{LoS}}(d)= (1-p_{\mathrm{out}}(d))e^{-a_{\mathrm{LoS}}d}
\\  
p_{\mathrm{NLoS}}(d)= 1-p_{\mathrm{out}}(d)-p_{\mathrm{LoS}}(d),
\end{matrix}
\ee
where $a_{\mathrm{out}}=0.0334$ m$^{-1}$, $b_{\mathrm{out}}=5.2$ and $a_{\mathrm{LoS}}=0.0149$ m$^{-1}$ (all these values are taken from~\cite{mustafa}). 
The long-term pathloss, which includes distance dependent attenuation and log-normal shadowing, is then given by:
\be
PL(d)[dB] = \alpha + \beta 10 \log_{10}(d) + \xi,
\ee
where $\xi\sim N(0,\sigma^2)$ is the log-normal shadowing, and parameters $\alpha,\beta,\sigma$ are reported in~\cite{mustafa} for both the 28 and the 73 GHz bands, and for both LoS and NLoS.

\section {Hybrid UE association}
\label{Hybrid_UE_association}
Simulations are run in an area $A$ where users and base stations are placed following a PPP deployment for each mobile network $m \in \mathcal{M}$.
In the primary phase of the algorithm, each UE $j \in \mathcal{J}_m$ in the area is associated to the BS $i \in \mathcal{I}_m$ that provides the minimum pathloss.
More precisely, the pathloss is computed under the same state condition in both bands.
In practice this means that if the link between a UE and a BS is in one of the three states (LoS, NLoS or outage), this state will be the same for both the 28 and 73~GHz bands. 
After the selection of the best BS, we randomly associate the UE to the BS band at 28 or 73 GHz, according to the probabilities $P_{28}$ and $P_{73} = 1 - P_{28}$.
In this study, these probabilities are taken equal to $\tfrac{1}{2}$ (we have verified by simulation that in most cases users are roughly equally split between the two bands).

In the second and last step, we associate a reference UE $j^*$, located at the center of the area, to the BS $i^*$ and carrier $c^*$ that maximize the throughput. 
The optimization problem can be formulated as follows:
\begin{equation}
\left( c^{*}, i^{*} \right)= \argmax_{i \in \mathcal{I}_m, c \in \mathcal {C}} \left( \frac{BW^{(c)}}{1 + N^{(c)}_{mi}} \log_2\left( 1 + \gamma_{mij^*}^{(c)}  \right) \right),
\label{max_th_equ}
\end{equation}
where $N_{mi}^{(c)}$ is the number of users already associated to the $c$-th carrier of the $i$-th base station and $BW^{(c)}$ is the bandwidth of the $c$-th carrier, $c \in  \mathcal{C} = \{28~\text{GHz}, 73~\text{GHz}\}$.
Then, we define the Signal-to-Interference-plus-Noise Ratio (SINR) term $\gamma_{mij^*}^{(c)}$ that considers the instantaneous interference value for UE $j^*$ allocated to the $c$-th carrier of the $i$-th BS, as:
\begin{equation}
\gamma_{mij^*}^{(c)} = \frac{\frac{P^{(c)}_{\text{TX}_{mi}}}{PL_{mij^*}^{(c)}}G^{(c)}_{mij^*}}
{\sum_{k \neq i}  \frac{P^{(c)}_{\text{TX}_{mk}}}{PL^{(c)}_{mkj^*}}G^{(c)}_{mkj^*} + BW^{(c)}  N_0},
\label{equation_sinr}
\end{equation}
where $P^{(c)}_{\text{TX}_{mi}}$ denotes the transmit power, $G^{(c)}_{mij^*}$ denotes the antenna gain at the $c$-th carrier (computed in the following), $k$ represents each interfering link, $G_{mkj^*}^{(c)}$ is the antenna gain between BS ${k \neq i}$ and UE $j^*$, and $N_0$ is the thermal noise. 
Then, $PL_{mij^*}^{(c)}$ denotes the pathloss between UE $j^*$ and the $c$-th carrier of the $i$-th BS. 
We assume a perfect alignment between the BS and UE beams so as to achieve the maximum beamforming gain  $G\, [\text{dB}] =  10 \log_{10} (n_{\text{TX}} n_{\text{RX}})$.
Under the assumption of perfect alignment between BS and UE beams, the antenna gain of the intended link is only a function of the carrier used (i.e., at 73~GHz there is a higher gain than at 28~GHz thanks to the higher number of antennas).
Therefore, the maximum beamforming gains $G^{(c)}_{mij^*}$ along the aligned directions, at 28~GHz and 73~GHz, are:
\begin{equation}
\centering
\begin{split}
G^{(28\, \text{GHz})} &= 10 \log_{10} (64 \cdot 16) \simeq  30\, \text{dB}\\
G^{(73\, \text{GHz})} &= 10 \log_{10} (265 \cdot 64) \simeq  42\, \text{dB}.
\label{max_gains}
\end{split}
\end{equation}
It is important to highlight that $G^{(c)}_{mij^*}$ is computed as in Equation~\eqref{bf_comp}, where the actual antenna gain will be affected by the instantaneous channel fading.
In a different manner, we note that the antenna gain $G_{mkj^*}^{(c)}$ between BS ${k \neq i}$ and UE $j^*$ is calculated assuming there is no alignment between transmit and receive beams, thus following Equation~\eqref{bf_comp} with a realistic pattern and angle that depends on the relative positions of UE $j^*$ and interfering BS $k$ for operator $m$. In other words, we assume beam alignment for the link between a UE and its serving BS, while we do not assume beam alignment between a UE and an interfering BS.

Using the instantaneous SINR in (\ref{equation_sinr}), we can derive the throughput of the reference user $j^*$ as:
\begin{equation}
\eta_{mj^*} = \frac{BW^{(c^*)}}{1 + N_{mi^*}^{(c^*)}} \log_2\left( 1 + \gamma_{mi^*j^*}^{(c^*)} \right).
\end{equation}
Repeating this two-step procedure a sufficient number of times (10$^4$ in our results), we are able to capture the statistics of interest for the reference user and to evaluate the performance of the hybrid association procedure.   

We summarize the proposed hybrid UE association procedure in Algorithm~\ref{association_procedure}.
\begin{algorithm}
\caption{}
\label{association_procedure}
\begin{algorithmic}[1]
\State $\forall m \in \mathcal{M}$ deploy in the area $A$ $J_m$ UEs and $I_m$ BSs following a PPP;
\For {$\forall$ operator $m \in \mathcal{M}$}
\For {$\forall$ user $j \in \mathcal{J}_m$}
\For {$\forall$ BS $i \in \mathcal{I}_m$}
\State $PL_{mij} \gets$ compute pathloss $\forall$ link $(i,j)$;
\EndFor
\EndFor
\EndFor
\State $N$: matrix initialized to zeros used to count \# UEs $\forall i \in \mathcal{I}_m$ and $\forall c \in \mathcal{C}$;
\State $\overline{M}$: vector that stores for each UE the $index$ of the associated BS;
\For {$\forall$ user $j \in \mathcal{J}_m$ and $\forall m \in \mathcal{M}$}
\State Associate user $j$ to the BS $i^*$ with minimum $PL_{mij}$;
\State $p \gets$ randomly pick a value $\in [0,1]$;
\If {$p < 0.5$}
\State $c^{*} \gets$ 28~GHz band;
\Else
\State $c^{*} \gets$ 73~GHz band;
\EndIf 
\State $N(i^{*}, c^{*}) \gets N(i^{*}, c^{*})  +1$;
\State $\overline{M}(j) \gets i^{*}$;
\EndFor
\State $P_{\text{TX}_{ic}}$: set equal to 30~dBm $\forall i \in \mathcal{I}_m$ and $\forall c \in \mathcal{C}$;
\State $G_{c}$: computed following Equations~\eqref{max_gains} $\forall c \in \mathcal{C}$; 
\State $BW_{c}$: bandwidth set $\forall c \in \mathcal{C}$;
\For {$\forall$ BS $i \in \mathcal{I}_m$}
\State $\overline{\gamma}_{mij^*}^{(c)} \gets$ compute matrix of SINRs $\forall i \in \mathcal{I}_m$, $\forall c \in \mathcal{C}$ as in~\eqref{equation_sinr};
\State $\eta_{mj^*} \gets$ compute matrix of rates using $\overline{\gamma}_{mij^*}^{(c)}$, $BW_c$, and $N(i,c)$;
\State $\left( c^{*}, i^{*} \right) \gets \argmax_{i \in \mathcal{I}_m, c \in \mathcal{C}}{(\eta_{mj^*})}$;
\EndFor
\end{algorithmic}
\end{algorithm}
\vspace{1cm}

We note that:
\begin{figure*}[t!]
\centering
\includegraphics[width=0.8\textwidth]{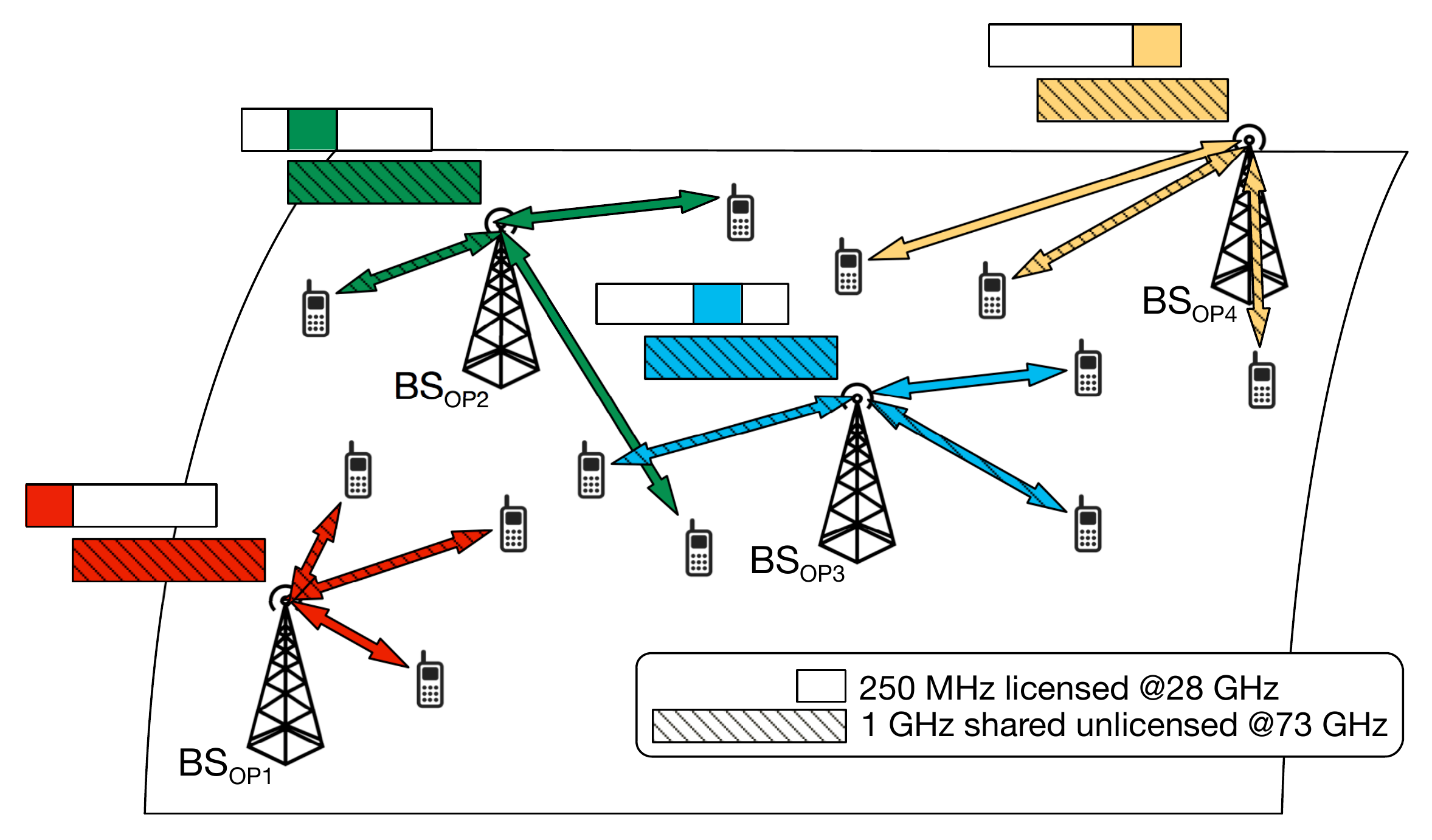}     
\caption{Hybrid spectrum access scheme. The 1~GHz band at 28~GHz is split among the four operators (exclusive spectrum access), while at 73~GHz the entire band of 1~GHz is shared among the operators (pooling).}
\label{scenario_draw}
\vspace{-0.65cm}
\end{figure*}
\begin{itemize}
\item The approach in (\ref{max_th_equ}) maximizes the throughput of the reference UE after association of all the previous users according to a minimum pathloss criterion and to a random decision for the carrier. Therefore, this procedure leads to a suboptimal result. However, due to the large number of BSs and UEs in the system, we believe this proposal provides a good starting point to analyze the performance of hybrid spectrum allocation. We plan to study the gap with respect to the optimal allocation in our future work.

\item The objective function in (\ref{max_th_equ}) considers the UE throughput. Another possible approach would be to consider the aggregate throughput within each BS or the total throughput in the system. We plan to explore these different approaches in our future work.

\item We are currently assuming that the allocation algorithm is implemented in a centralized way. However, the algorithm could be implemented in a distributed way, assuming that a supporting signaling between BS and UE and inter-operator interference estimation capabilities at the UEs are available. 

 \item The  approach in (\ref{max_th_equ}) assumes full buffer UEs and round-robin scheduling. However, the results can be extended to a proportionally fair scheduler by considering also the effect of a different required rate for each UE. 
\end{itemize}

\section{Simulation results}
\label{simulation_results}
In this section, we compare the results obtained through our proposed hybrid spectrum access, as depicted in Figure~\ref{scenario_draw}, against two baselines: one adopts exclusive spectrum access at both 28~GHz and 73~GHz and the other uses full spectrum pooling at both carriers.

We performed simulations also varying the number of antenna elements for the 73~GHz band.
More precisely, we simulated the following two configurations:
\begin{description}
\item [\emph{i)}] Both bands use the same number of antenna elements $n_{\text{TX}} = 64$ and  $n_{\text{RX}} = 16$.
\item [\emph{ii)}] We double the number of antenna elements per dimension for the 73~GHz band, i.e., $n_{\text{TX}} = 256$ and  $n_{\text{RX}} = 64$, while keeping the configuration at 28~GHz as in case \emph{i)}.
\end{description} 

\begin{figure}[t!]
\centering
\includegraphics[width=\columnwidth]{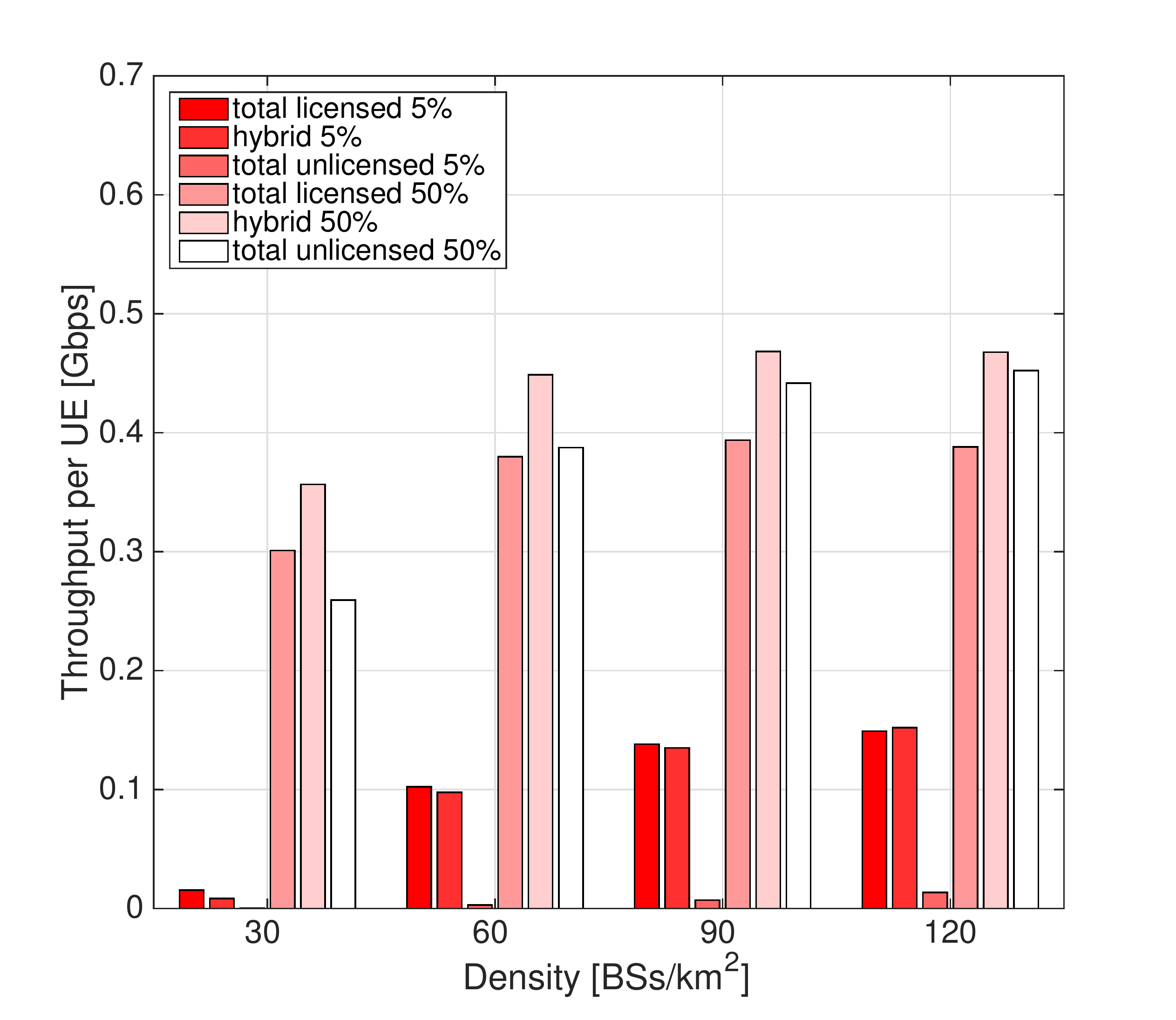}     
\caption{Comparison of the average throughput $\eta$ measured for the hybrid case and the two baselines. Values for the 5-th and 50-th percentiles in case \emph{i)}, where both bands use the same number of antenna elements.}
\label{case_ii} 
\end{figure}

\begin{figure}[t!]
\centering
\includegraphics[width=\columnwidth]{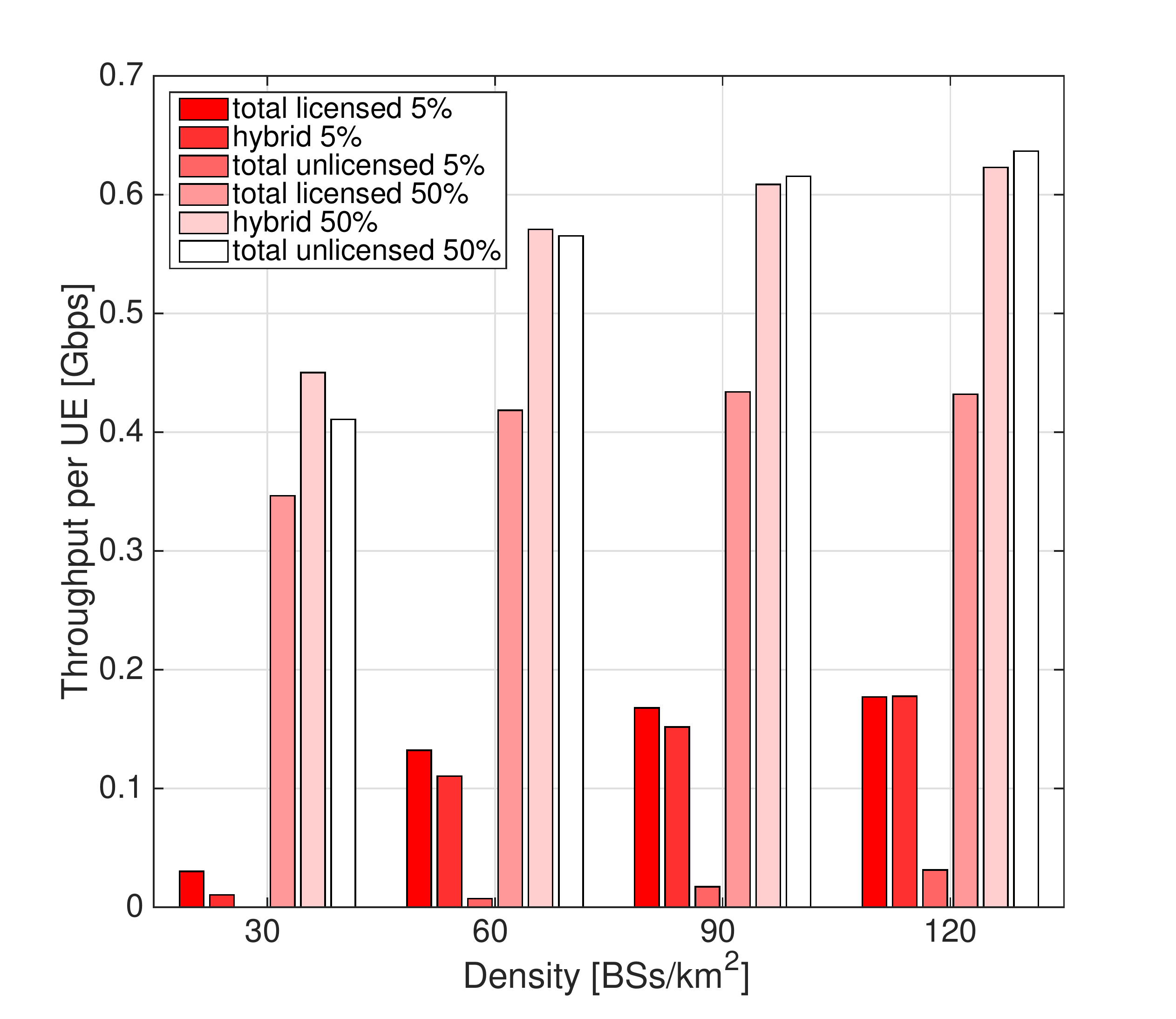}     
\caption{Comparison of the average throughput $\eta$ measured for the hybrid case and the two baselines. Values for the 5-th and 50-th percentiles in case \emph{ii)}, where the carriers use a different number of antenna elements.}
\label{case_iii} 
\end{figure}

As a result, we show in Figures~\ref{case_ii} and \ref{case_iii} the average throughput measured for each configuration.
If the band is \emph{licensed}, each operator has its own spectrum, orthogonal to the others, which results in 250~MHz of available bandwidth. 
Conversely, if the band is \emph{unlicensed}, all the operators in the area share the entire spectrum, which results in 1~GHz of available bandwidth. 

\smallskip
\textbf{More antennas $\rightarrow$ higher throughput}
\smallskip

By comparing Figures~\ref{case_ii} and \ref{case_iii}, we can observe a first, expected result: in case \emph{ii)} -- depicted in Figure~\ref{case_iii} -- which corresponds to a scenario where we deploy more antennas at both TX and RX for the 73~GHz carrier, the average throughput per UE increases, for both 5-th percentile and median user. This is motivated by the increased antenna gain obtained by adopting a higher number of antennas. 

\smallskip
On the other hand, two extremely insightful and promising trends are captured by the following considerations, which fully validate our \emph{hybrid-access} intuition.

\smallskip
\textbf{Hybrid access vs. fully licensed}
\smallskip

The performance ratio of the two licensing schemes increases with the BS density, in favor of our proposed hybrid access. This promising result holds for both the worst-case (5\%) and the median (50\%) users, and for both system configurations, i.e., case \emph{i)} and case \emph{ii)}, as shown in Figures~\ref{case_ii} and~\ref{case_iii}, respectively. In other words, we observe increasing gains when we enable the option to opportunistically choose between two carriers, which allows to intrinsically capture both interference and available bandwidth. 

\smallskip
\textbf{Hybrid access vs. fully unlicensed}
\smallskip

On the other hand, we are interested in comparing the performance trends of a hybrid access scheme against that of a fully unlicensed approach. We note that the median user (50\%) throughput when adopting a fully unlicensed access strategy increases with the BS density; the gap with respect to the performance obtained through a hybrid scheme reduces until (i) almost overlapping, in Figure~\ref{case_ii}, and (ii) slightly outperforming it, in Figure~\ref{case_iii}. However, if we observe the performance trends of the worst-case users (5\%), our proposed hybrid access strategy always greatly outperforms the achieved throughput of a fully unlicensed policy. 

\smallskip

This preliminary set of results shows that a hybrid allocation procedure can provide a promising tradeoff between fully licensed and fully unlicensed policies. In particular, it can be observed that, for the median user, in most cases (except for low density in case \emph{i)}) the fully unlicensed policy has better performance than the fully licensed one. The opposite is true for the worst-case users, i.e., the fully unlicensed solution is always much worse that the fully licensed one. 
On the other hand, our proposed hybrid policy often performs better than both baselines, or it outperforms one and is comparable (or only slightly inferior) to the other, and is therefore able to provide consistently good performance across all users.

\section{Conclusion}
\label{conclusion}
We  studied the performance of a hybrid spectrum scheme where exclusive access is used at frequencies in the 20/30~GHz range while spectrum pooling/unlicensed spectrum is used at frequencies around 70~GHz.
Our preliminary results show that hybrid spectrum access is a promising approach for mmWave networks and motivate further studies to achieve a more complete understanding of both technical and non technical implications.

\bibliographystyle{IEEEtran}
\bibliography{biblio}

\end{document}